\pdfoutput=1
\documentclass[aps,twocolumn, floatfix, prb]{revtex4}
\usepackage{graphicx}
\DeclareGraphicsExtensions{.pdf}
\usepackage{amsmath,amssymb,bbold,bm}
\usepackage{float}
\usepackage{epstopdf}

\newcommand{\bk}{{\bm k}}
\newcommand{\br}{{\bm r}}
\newcommand{\bR}{{\bm R}}
\newcommand{\bB}{{\bf B}}
\newcommand{\bA}{{\bm A}}

\newcommand{\bs}{{\bm s}}

\def\sgn{\mathop{\rm sgn}\nolimits}

\begin{document}

\title{Spin Response of Electrons on the Surface of a Topological Insulator}

\author{M.M. Vazifeh}
\author{M. Franz}
\affiliation{Department of Physics and Astronomy, University of
British Columbia, Vancouver, BC, Canada V6T 1Z1}

\begin{abstract}
The surface of a topological insulator hosts a very special form of a quasi-two dimensional metallic system when 
it is embedded in a topologically trivial medium like the vacuum. The electronic properties of this unusual
2D metal are distinct in many aspects from both the conventional two-dimensional electron gas systems in quantum 
well heterostructures as well as those of a single layer graphene. In this paper, we study one of these distinct 
features i.e., the response of the electronic spins to an applied magnetic field perpendicular to the surface. 
We find an unusual behaviour of the spin magnetization and
susceptibility as a function of both the magnetic field and the
chemical potential for a generic topological surface.  We propose that
this behavior could be studied by the recently developed experimental 
 technique called $\beta$-NMR which is highly sensitive to the surface
 electron spins. We explain how this technique could be used to probe for spontaneous
magnetic ordering  caused by magnetic dopants or interactions
discussed in the recent literature.
\end{abstract}

\date{\today}

\maketitle

\section{Introduction}
Topologically protected electronic states residing on the surface of topological 
insulators (TI) \cite{hasan1, hasan2, shen1,hasan3, hasan4, fu-kane3D,
  hasan_rev, Zhanybek} form a unique 2D metal distinct from
those so far realized in solid state systems. Similar to the 
low energy electronic states in a single layer graphene\cite{Neto},
the robust 2D metal on an TI surface has 
conic branches touching (in the absence of the intrinsic gap) at
high-symmetry points in the first Brillouin zone. 
This resemblance in the energy dispersion is responsible for some common properties between these metallic systems, such as the square-root dependence of Landau Levels (LLs) to the applied magnetic 
field. However, there are important differences between these systems
as a result of the topological nature of the surface states in TIs.
The low energy Dirac-like surface bands of a topological insulator
arise due to the mismatch of the bulk topological invariants 
on the two sides of the surface regardless of how constituent atoms
have been arranged on the surface as long as the bulk remains in the 
topological phase\cite{fu-kane3D,zhangprb}. In the case of a `strong'
topological insulator (STI)  there is an odd number of Dirac points at
any surface and, importantly, they are not spin
degenerate. Furthermore, they exhibit  a unique spin-momentum 
locking\cite{hasan3, hasan4} as a result of the strong spin-orbit
interaction in the underlying STI which is 
essential for the formation of the topologically non-trivial insulator
phase with time-reversal symmetry. These 
properties make electronic surface states in a STI very distinct from
those of graphene with a pair of spin 
degenerate Dirac-like bands near two special points in the Brillouin
zone which exist as a result of the
unique arrangement of carbon atoms in the honeycomb lattice.
One of the interesting aspects in which these 2D metallic systems
behave uniquely is their magnetic response.\cite{DiracMag2} 
Here, motivated by the recent progres in the experimental methods and 
the importance of the spin susceptibility for our understanding of electronic
systems, we study the spin response of electrons on the surface of a
STI and show that it exhibits interesting features even in a simple non-interacting limit. We find that the characteristic
spin-momentum locking of these electrons leads to a unique spin
response that is distinct from spin-degenerate systems like graphene
and 2DEG. Instead 
of the oscillatory behaviour of the susceptibility as a function of
the chemical potential found in spin degenerate systems, in STIs our
work predicts a plateau-type behaviour  which arises from the strong
correlation between spin and orbital degrees of freedom. In addition
we find that the existence of a special LL with full spin polarization
leads to a jump in the magnetization as the chemical potential crosses the 
energy of this LL. When an intrinsic magnetic ordering is present
there is also a jump in the magnetization as a function 
of the applied magnetic field. Our results can therefore assist in
detection of such intrinsic magnetization\cite{MagDopped, MagDopped2}
which is known to have profound consequences for the nature of the
surface state; the magnetized surface state of a TI is predicted to become a quantum Hall liquid with half-integer quantized Hall
conductivity\cite{zhangprb,fu2} and many
unusual\cite{dhlee1,qi2,wormhole,tse1,maciejko1,garate1} and potentially useful\cite{nagaosa1,garate2} physical properties.

There are
various ways of measuring the weak magnetization produced by electrons 
on a metallic surface. The SQUID scanning magnetometry is a highly
sensitive probe  which can detect tiny 
magnetization on the surfaces. However, it is a challenge to use this
device to probe magnetization when  there is a large applied magnetic
field which interferes with the superconducting part of the SQUID and causes
noise. A variant of this method using a superconducting pickup coil
has been developed in order to study 
deHaas-van Alphen oscillations of the 2D metallic systems in a
large perpendicular magnetic field.\cite{Squid1, Squid2} 
High-sensitity micro-mechanical cantilever magnetometry \cite{MagExp1}
is another way of measuring electronic magnetization, however, this
method measures the magnetization of the whole sample and it is difficult to isolate 
the contribution from the electrons on a single surface.    
The nuclear magnetic resonance (NMR) is another powerfull experimental
technique which can be used to study electronic spin magnetic response
in a bulk metal.\cite{nmrbook} Through the so-called Knight shift in the 
nuclear resonance peak, it is possible to probe the spin part of the
magnetic susceptibility of the electrons as they interact with the
resonating nuclei in their proximity. Unfortunately, this method, in
its conventional form, fails to be useful in very thin films and 2D
metallic systems due to the limitation in the number of available nuclei and 
the resulting weakness of the signal. Thanks to the progresses made by 
experimentalists in controlling and implementing 
high energy beams of unstable ions, an experimental technique, the
so-called $\beta$-NMR, has been recently developed to overcome the
above limitations. Briefly, in this exotic variety of NMR, unstable
radioactive ions such as $^8$Li and $^{11}$Be are implanted in the
sample surface. The nuclear spin precession signal is then detected
through the products of the beta decay of the radioactive nucleus.
Since the ion implantation depth can be controlled by tuning the beam
energy it is possible to acquire information about the behaviour of the 
electronic spins in very thin metallic films.\cite{betaNMR1,betaNMR2}
Experiments are currently underway to study the surface magnetic response
of STI crystals Bi$_2$Se$_3$ and Bi$_2$Te$_3$ using
$\beta$-NMR.\cite{farlane}  
The paper is organized as follows. First we introduce a simple model
known to describe the electrons on the surface of 
a STI in the presence of a perpendicular magnetic field. We review the
exact solutions of its eigenvalue problem which has been studied
previously in various contexts.\cite{DiracMag1,DiracMag2} In section
III, we calculate the spin magnetization and susceptibility assuming
that the chemical potential lies inside the gap between positive and
negative energy eigenstates. We discuss the magnetic response of the
surface as one tunes the magnetic field and the chemical potential for 
various values of the intrinsic gap.  In closing, we explain how a $\beta$-NMR experiment might be able to detect these effects through Knight shift measurements.

\section{The Model}

Many of the interesting features of the surface states in a topological insulator can be captured, at least qualitatively, 
using a simple non-interacting Dirac Hamiltonian
\begin{equation}
H = \sum_{\bk}\Psi^{\dag}_{\bk} [\hbar \upsilon_F{\bm  \sigma} \cdot ( \hat{z} \times \bk) + \Delta_0 \sigma_z] \Psi_{\bk},
\end{equation}  
where $\Psi^{\dag}_{\bk} = (c^{\dag}_{\tiny \uparrow \bk}, c^{\dag}_{\tiny \downarrow \bk})$ and
$c^{\dag}_{\tiny \uparrow (\downarrow)\bk}$ is the fermionic creation
operator of the spin up(down) states with wave vector
$\bk$. $\Delta_0$ is the intrinsic gap in the surface spectrum which
might be nonzero when the time-reversal symmetry is spontaneously broken due to
magnetic ordering. The latter can arise due to the presence of
magnetic dopants with spin  ${\bm S}$ in the proximity of the surface
exchange-coupled to the electronic spins\cite{MagDopped,
  MagDopped2,Rosenberg} or as a result of electron-electron interaction\cite{stern1}.

Turning on a perpendicular magnetic field adds two terms 
to the above Hamiltonian. One is the minimal coupling of the magnetic
vector potential, $\hbar\bk \rightarrow \hbar\bk + e \bA$ (electron
charge $-e$). The other is the coupling of the spins to the
magnetic field, the Zeeman effect, expressed as
$\delta H_Z= - g_s\mu_B \hbar^{-1} \bB\cdot\bs$, where $g_s$ is the 
effective electron gyromagnetic constant.  In the bulk  Bi$_2$Se$_3$ crystal  $g_s\simeq 30$ (Ref.\ \onlinecite{ZhangFP}) although much smaller values have been reported for electrons near the surface.\cite{ong1} In our calculations below we use two representative values, $g_s=8$ and $g_s=30$, which yield qualitatively similar results with some interesting differences.
In the continuum limit, the leading order Hamiltonian describing these
surface states in the presence of the 
applied magnetic field, $\bB = B_0 \hat{z}$, in the Landau gauge $\bA = -(B_0 y, 0)$, can be written in the 
following form
\begin{equation}
H =  \sum_{k_x} \int dy \;\Psi^{\dag}_{k_x}(y) {\cal H}(k_x,y)\Psi_{k_x}(y),
\end{equation}
where $\Psi^{\dag}_{k_x}(y)$ is the creation operator for the spinor
mode extended along the $x$ direction and 
localized at $y$.  ${\cal H}(k_x,y)$ for $B_0>0$ is defined as 
%
\begin{figure}[t]
\includegraphics[width=8cm]{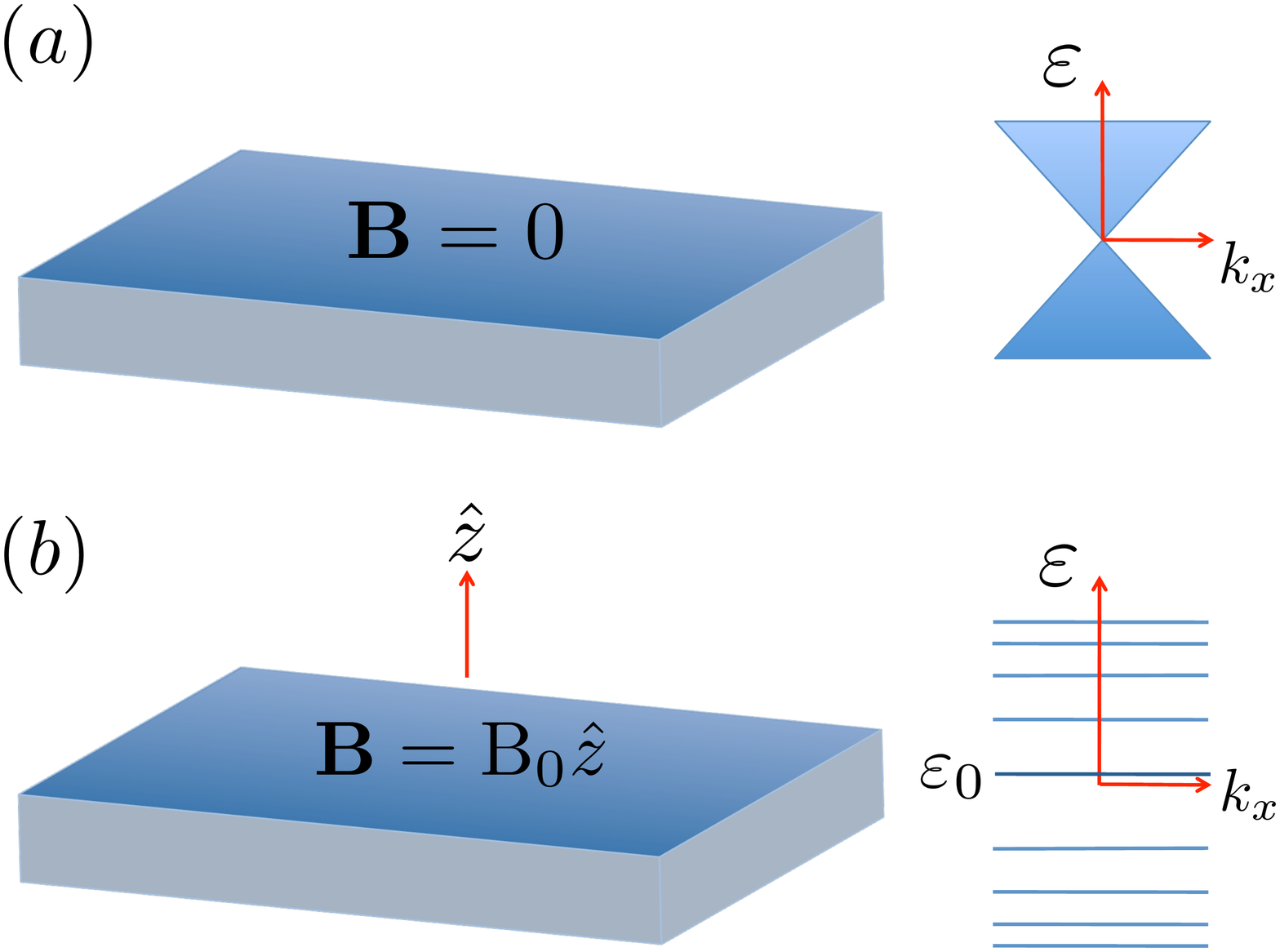}
\caption{(Color online) The surface spectrum of a strong topological insulator in the absence of magnetic dopants ($\Delta_0=0$), (a) in the absence of the external magnetic field, (b) in the presence of an applied perpendicular magnetic field.}\label{fig1}
\end{figure}
%
\begin{equation}\label{H1}
{\cal H}(k_x,y) = \left(\begin{matrix}  \Delta & i \epsilon_c a_{k_x} \\ -i \epsilon_c a^{\dag}_{k_x} & -\Delta \end{matrix}\right),
\end{equation}
where  $\epsilon_c = \upsilon_F \sqrt{2 e \hbar |B_0|}$ and 
\begin{equation}\label{del1}
\Delta= \Delta_0 - {g_s \mu_B B_0\over 2}.
\end{equation}
The term in $\Delta$ proportional to the magnetic field is the Zeeman contribution. 
 $a_{k_x}$ is the one-dimensional harmonic oscillator bosonic operator
 defined as
 \begin{equation}
a_{k_x} = {1 \over \sqrt{2}} \left(\frac{y}{l_B} + l_B (\partial_y- k_x)\right), \;\;\;\; l_B^2 =   {\hbar \over e |B_0|}.
\end{equation}
Note that varying $k_x$ shifts the position of the localized state
produced by the application of $a^{\dag}_{k_x}$ on the vacuum state
along the $y$ direction. In the $B_0<0$ case, ${\cal H}(k_x,y) $ can
be obtained from the one given for $B_0>0$ in Eq. (\ref{H1}) by
exchanging the off-diagonal elements and replacing $k_x$ by its time 
reversed counterpart $-k_x$. 

The eigenstates for $B_0>0$ and $n>0$ are given by \cite{DiracMag1,DiracMag2}
\begin{equation}\label{phi_p}
\phi^{+}_{k_x,n}(y) =  \left(\begin{matrix}\cos{(\delta_n/2)}\varphi_{n-1}(y-k_x {l_B}^2) \\ \\ -i \sin{(\delta_n/2)}\varphi_{n}(y-k_x {l_B}^2)\end{matrix}\right),
\end{equation} 
\\
\begin{equation}\label{phi_m}
\phi^{-}_{k_x,n}(y) = \left(\begin{matrix}\sin{(\delta_n/2)}\varphi_{n-1}(y-k_x {l_B}^2) \\ \\ i \cos{(\delta_n/2)}\varphi_{n}(y-k_x {l_B}^2)\end{matrix}\right),
\end{equation} 
while for  $n=0$ we have 
\begin{equation}\label{phi0}
\phi_{k_x,0}(y) =  \left(\begin{matrix}0 \\ \\ \varphi_{0}(y-k_x {l_B}^2)\end{matrix}\right).
\end{equation} 
In the above $\cos{\delta_n} = {\Delta / \varepsilon^{+}_n}$ and
$\varphi_n$ are the one-dimensional 
harmonic oscillator eigenstates, i.e., $a^{\dag}a \varphi_n = n \varphi_n$.
We remark that the $n=0$ eigenstate in Eq.\
(\ref{phi0}) is very special since it is fully {\em spin
  polarized}. This will have important consequences for the magnetic
response discussed below.

The eigenvalues associated with $\phi^{\pm}_{k_x,n}(y)$ eigenstates are given by 
\begin{equation}
\varepsilon^{\pm}_n = \pm \sqrt{n \epsilon_c^2 + \Delta^2}, \;\;\;\;\;\; n>0  
\end{equation}
and for the fully spin-polarized $n=0$ eigenstates 
\begin{equation}
\varepsilon_0 = -\sgn{(B_0)} \Delta.
\end{equation}
Note that the form of eigenstates for $B_0<0$ is different from that given in Eq.(\ref{phi_p}-\ref{phi0}) 
since the Hamiltonian is different in that case.

\section{Spin Susceptibility and magnetization}

The electronic magnetic moment due to spin is proportional to the spin operator
\begin{equation}
{\bm \mu}_e = - \gamma_e {\bm s},  \;\;\;\;\;\;\;\;\; (\gamma_e = - g_s \mu_B / \hbar)
\end{equation}
Therefore, to calculate the spin part of the magnetic moment for the eigenstates given in the previous section 
we only need to find the expectation value of the spin operator. A
straightforward evaluation using Eqs.\ (\ref{phi_p}-\ref{phi0}) shows
that all the electronic states within  the same LL contribute equally
to the magnetization. For each of them we have 
\begin{center}
\begin{equation}
{\cal M}^{\alpha}_{x,n} = {\cal M}^{\alpha}_{y,n} = 0
\end{equation}
\begin{equation}\label{mzn}
{\cal M}^{\alpha}_{z,n}  =   \frac{g_s \mu_B}{2}\cdot\frac{\Delta}{\varepsilon^{\alpha}_n}
\end{equation}
\begin{equation}\label{mz0}
{\cal M}_{z,0} = - \frac{g_s \mu_B}{2} \sgn{(B_0)}
\end{equation}
\end{center}
%
\begin{figure}[t]
\includegraphics[width=8cm]{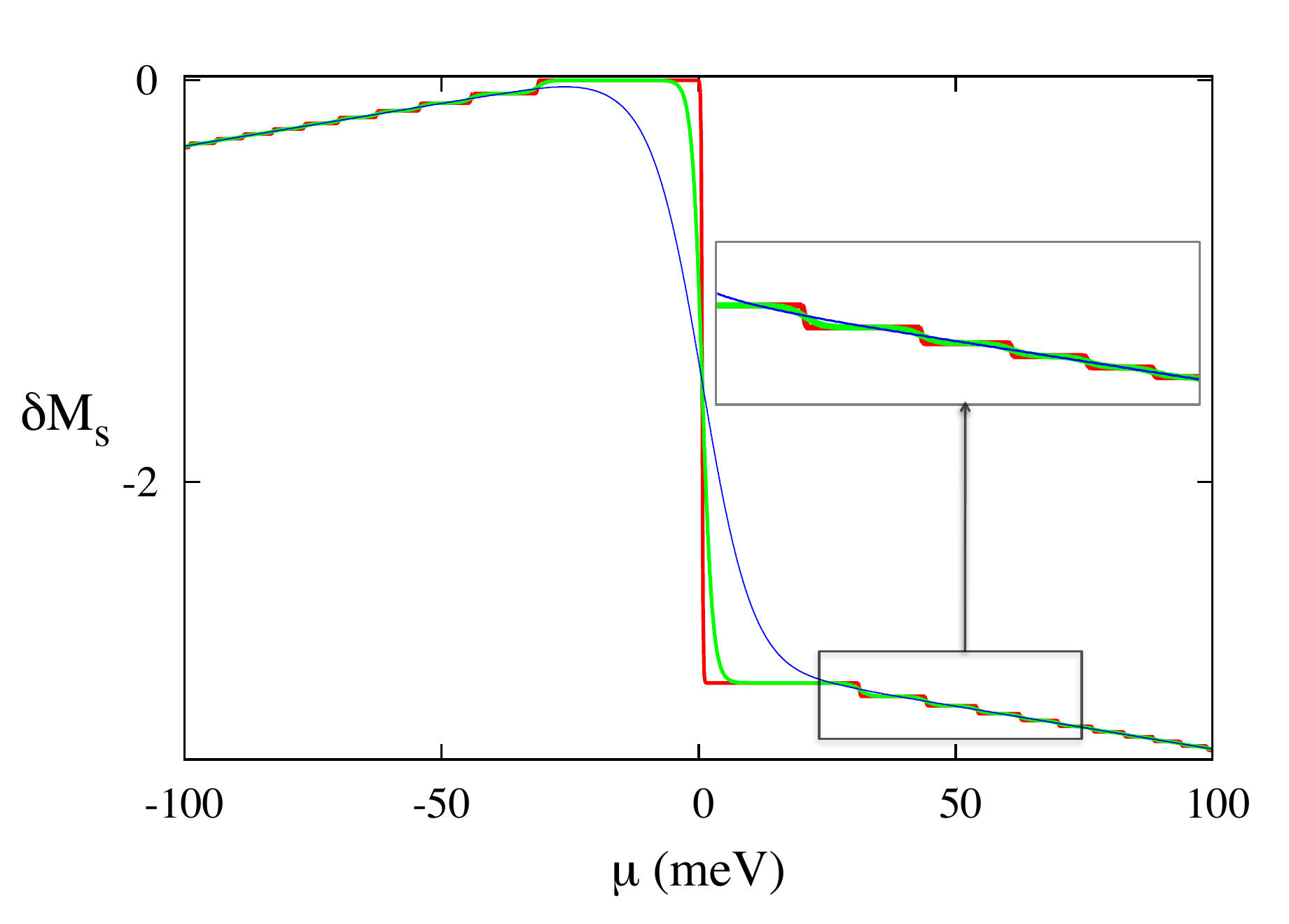}
\includegraphics[width=8cm]{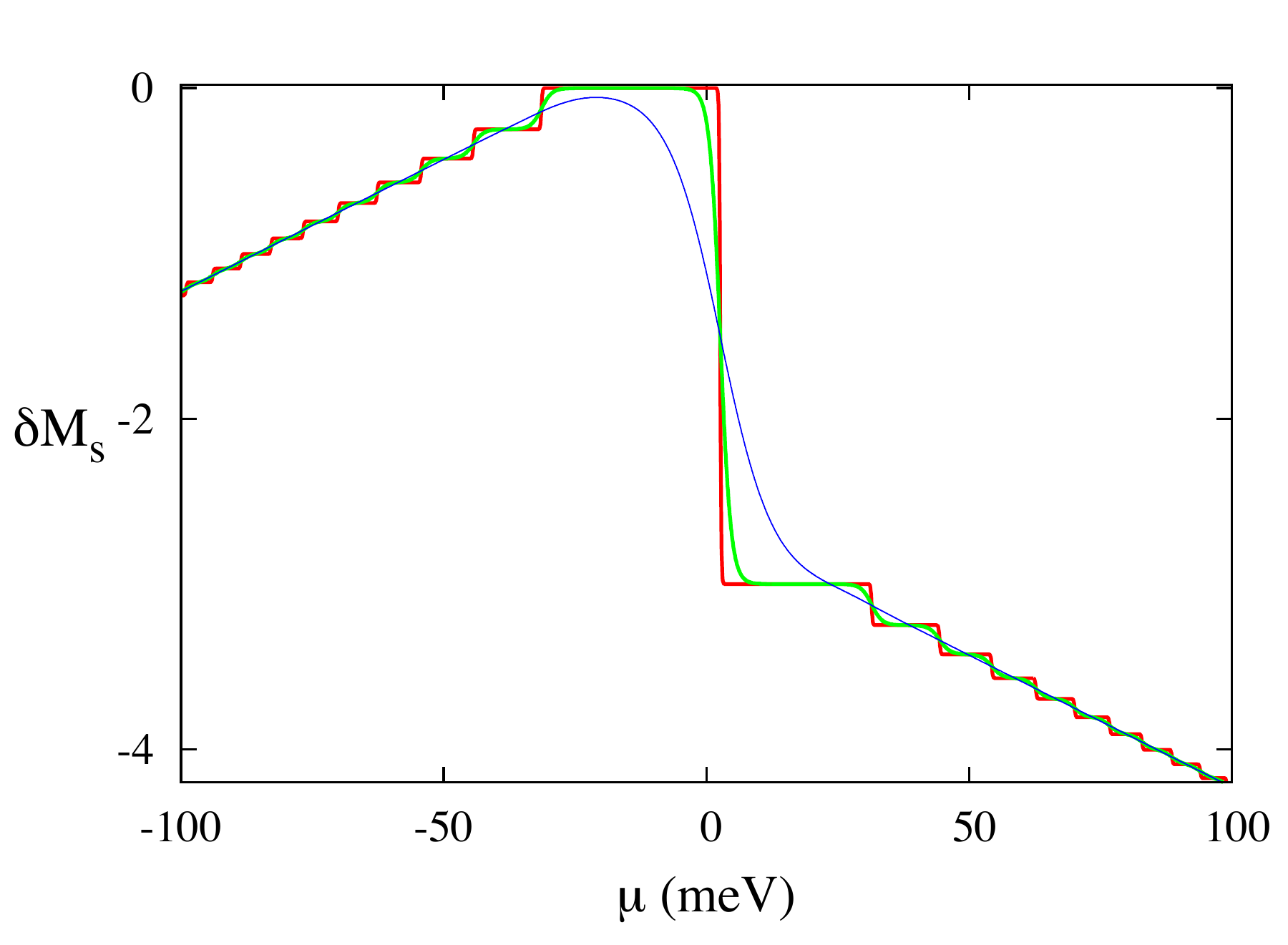}
\caption{(Color online) Spin magnetization $\delta M_s = M_s(T,\mu) - M_s (0,0)$  in units of ($\chi_0 \cdot$ Tesla) as a function of the 
chemical potential for a nonmagnetic surface ($\Delta_0 = 0$) at $\text{B}_0 = 3.0$ Tesla, $g_s = 8$ (top panel) and $30$ (bottom panel). $k_{\text{B}} T = 0.1,1.0,5.0$ meV 
(red, green, blue).}\label{fig2}
\end{figure}

The total magnetization for each Landau level can be obtained by
multiplying the above quantities by the Landau level 
degeneracy $L^2/(2 \pi l_B^2)$, representing the total number of
states with characteristic length $l_B$ that the surface 
area $L^2$ can accommodate. Since the magnetization contribution
computed above for each eigenstate is an 
explicit function of its energy, we can perform the following integral
to find the total magnetization density due to the  electronic spins
\begin{equation}\label{eq1}
M_s = \int d\varepsilon \; D(\varepsilon) {\cal M}_s(\varepsilon) n_F(\varepsilon),
\end{equation}
where ${\cal M}_s(\varepsilon)= g_s \mu_B \Delta/(2\varepsilon)$ is
the magnetization of the eigenstate with energy $\varepsilon$ and
$n_F(\varepsilon)=1/[e^{(\varepsilon-\mu)/k_BT}+1]$ is the Fermi-Dirac distribution function. The 
electronic density of states associated with the surface states is $D(\varepsilon)$. For the Hamiltonian we used in the previous section it takes the following form
\begin{equation}
D(\varepsilon) = \frac{1}{2\pi l_B^2} \left[\delta(\varepsilon- \varepsilon_0) + \sum_{{n>0,  \alpha=\pm}}^{n_c}\delta(\varepsilon- \varepsilon^{\alpha}_{n})  \right],
\end{equation}
where $n_c \equiv (\Lambda^2 - \Delta^2)/\epsilon_c^2$ is the Landau level index beyond which the 
energy exceeds the cutoff energy $\Lambda$. The cutoff can be chosen
to be the energy where the surface band becomes degenerate with the
bulk bands and here we assume $\Lambda=300$ meV. Using this density of states function to perform integration in Eq. ({\ref{eq1}}) yields
\begin{equation} \label{ms}
\frac{M_s}{B_0} =\chi_0 \left[ -n_F(\varepsilon_0) + \sgn{(B_0)} \sum_{n=1}^{n_c} \frac{n_F(\varepsilon^{+}_n)  - n_F(\varepsilon^{-}_n)}{\varepsilon^{+}_n/\Delta} \right],
\end{equation}
where $\chi_0 \equiv {(e g_s \mu_B)}/{ 2 h}$. 

\begin{figure}[t]
\includegraphics[width=8cm]{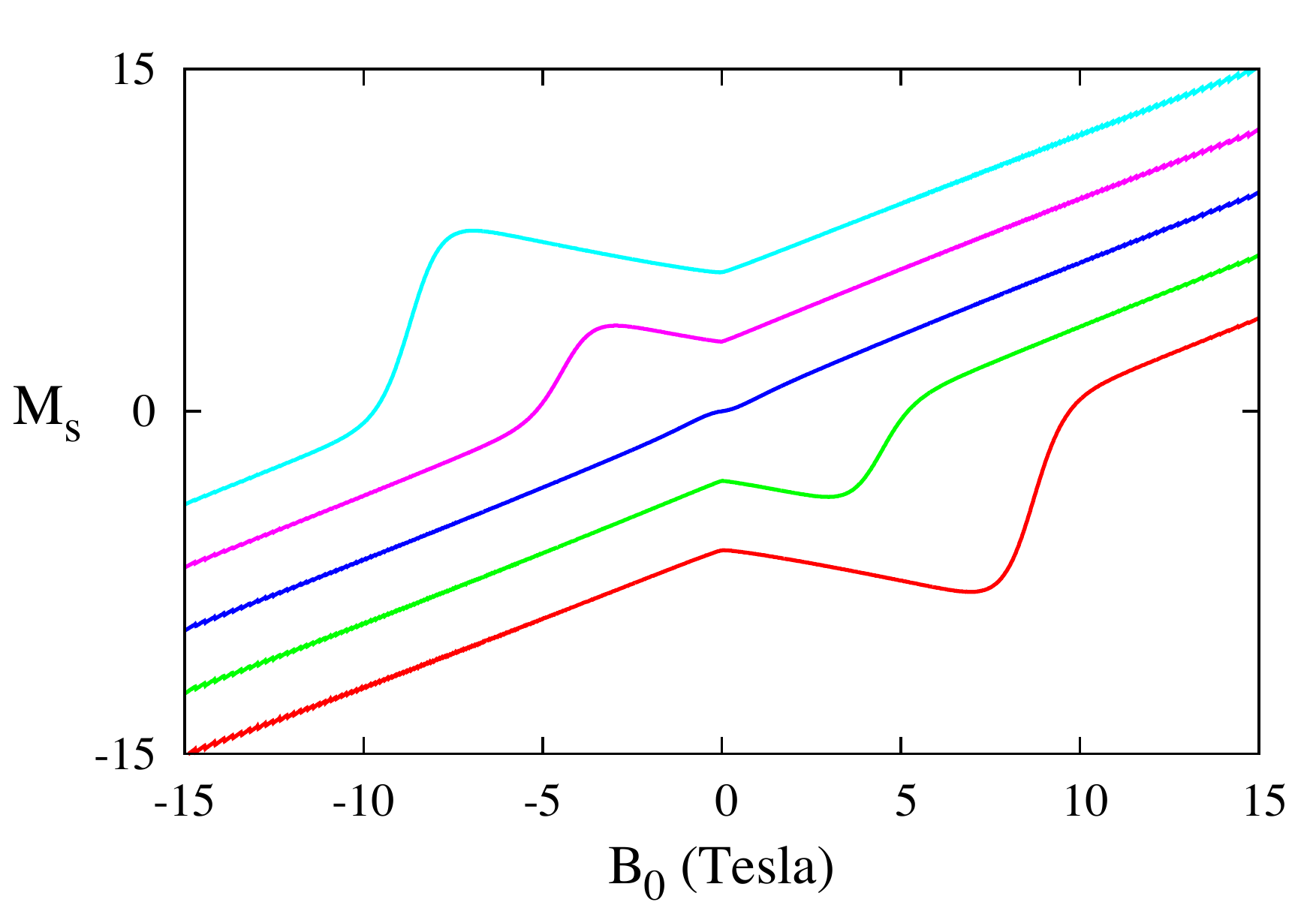}  
\includegraphics[width=8cm]{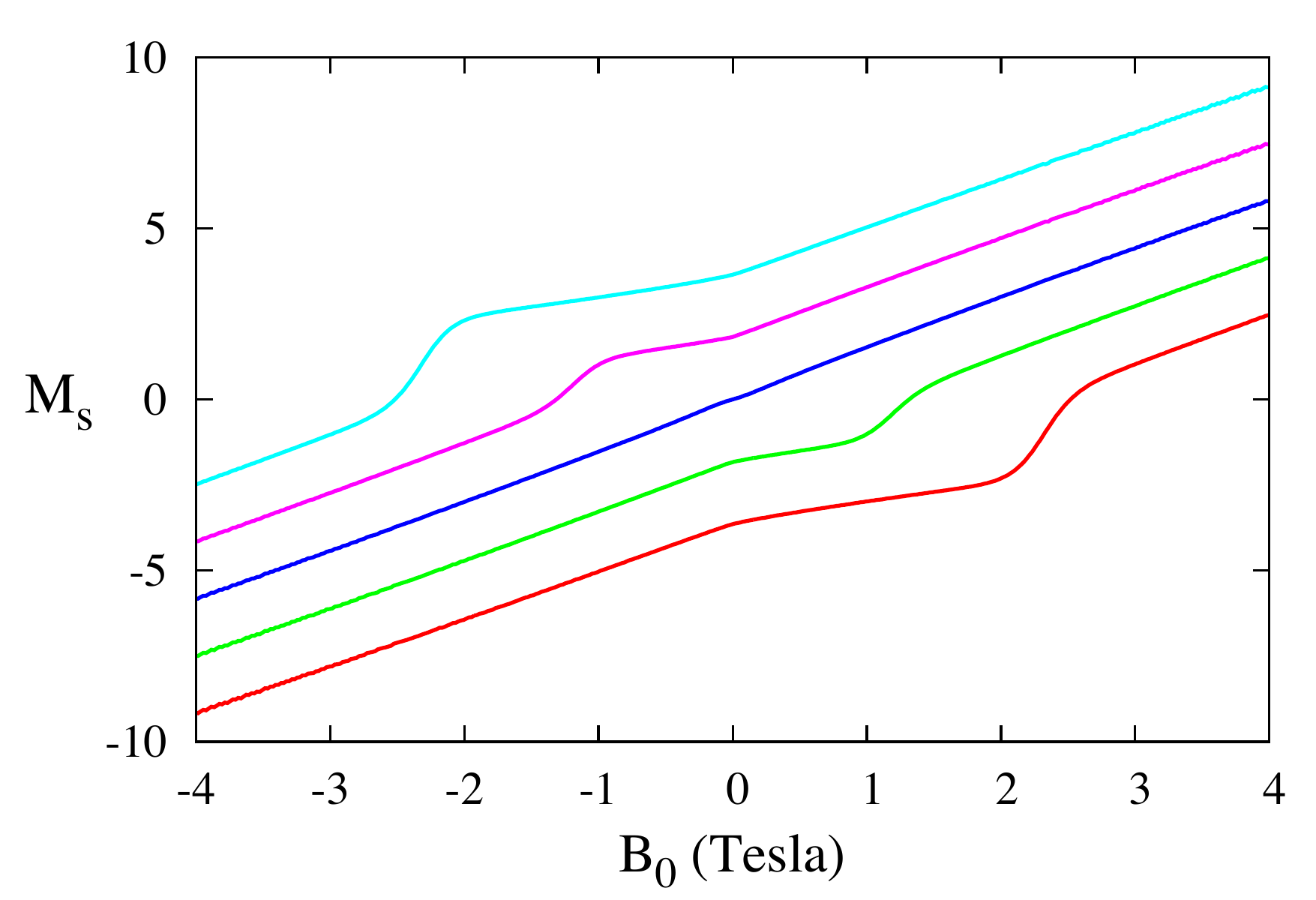}  
\caption{(Color online) Spin magnetization in units of ($\chi_0 \cdot$ Tesla) as a function of the magnetic field for 
(from left to right) $\Delta_0=-2,-1,0,1,2$ meV and $k_B T=0.1$meV. $g_s = $ 8 (top panel) and 30 (bottom panel). }
\end{figure}\label{fig3}

For a constant magnetic field $B_0$ the magnetization of a single TI
surface given by Eq.\ (\ref{ms}) shows an interesting behaviour as a
function of chemical potential $\mu$ illustrated in
Fig.\ \ref{fig2}. In order to avoid ambiguity associated with the
high-energy cutoff we choose to display $\delta M_s = M_s(T,\mu) - M_s
(0,0)$, i.e. spin magnetization relative to the neutrality point at $T=0$.
For the negative values of $\mu$ magnetization initially decreases
reflecting the fact that the negative-energy surface states exhibit
negative spin polarization, as can be seen from Eq.\ (\ref{mzn}). The
large jump in $\delta M_s$ near $\mu=0$ results from electrons filling
the fully spin-polarized $n=0$ Landau level. Further increase in $\mu$
results in increase of $\delta M_s$ now reflecting the fact that the 
positive-energy surface states exhibit positive spin polarization.

\begin{figure}[t]
\includegraphics[width=8cm]{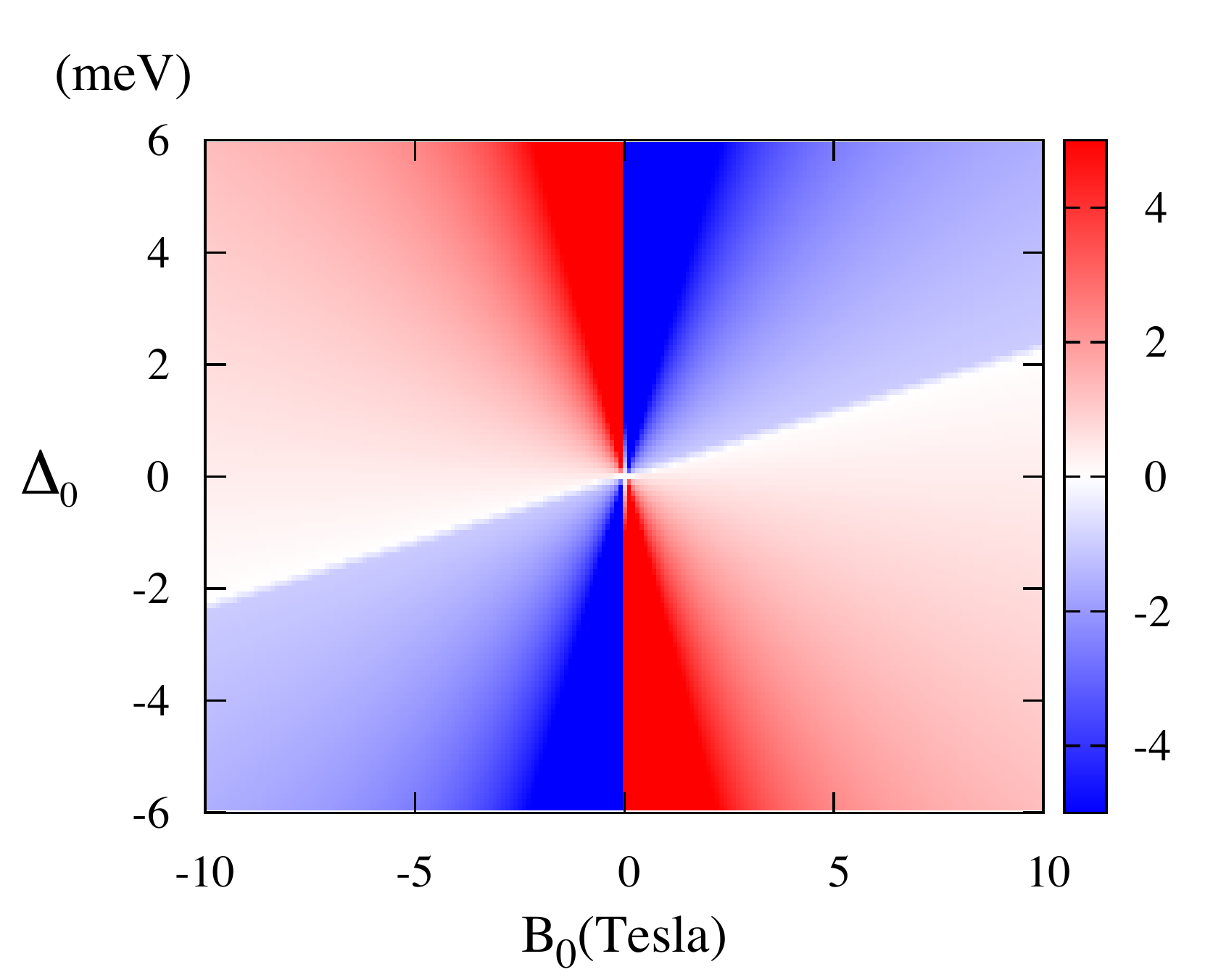}
\caption{(Color online) The color at each point $(B_0,\Delta_0)$ represents the 
magnitude of the spin susceptibility in the linear response regime, i.e., $\chi=M_s/B_0$, in units of $\chi_0$. The step (diagonal white line) is given by Eq. (\ref{gapeq}) and results from $n=0$ Landau lavel crossing the chemical potential. The discontinuity evident at $B_0=0$ reflects the fact that susceptibility $\chi$ diverges as $B_0\to 0$. We have assumed $g_s = 8$ and $k_B T = 0.01$meV in this graph. }\label{fig3}  
\end{figure}\label{fig4}
The above behavior is unique to topological insulators for it results
from the Landau level structure in a single Dirac point
(or more generally an odd number thereof). In a TI the `other' Dirac point is
located on the opposite surface where the magnetic field points in the
direction opposite relative to the surface normal, see Fig.\ 
\ref{fig1}. The contribution of
this surface to $\delta M_s$ would be the same. We emphasize that for
a relatively thick TI slab $\beta$-NMR will be sensitive to a single surface facing the beam
and the behaviour predicted here is in principle observable,
except that continuous tuning of the chemical potential in a crystal might be difficult to achieve.

A much more feasible experiment involves varying magnetic field $B_0$
while keeping $\mu$ constant. We now show that a unique signature of the
fully polarized $n=0$ Landau level still exists in samples with intrinsic magnetic ordering, when the chemical potential resides in the gap (i.e. both the bulk and the surface are insulating in the absence of the field). 
In his situation  we can set $\mu=0$ in our model. Now consider the effect of the applied magnetic field. Since the gap between the
adjacent energy levels in which the Fermi energy is located is fairly
large ($\sim 100$ K for a 1T field), at sufficiently low
temperatures we can replace the Fermi function in Eq. (\ref{ms}) by the step function  
$n_F(\varepsilon) \to \Theta(-\varepsilon)$ and write
\begin{equation} 
\frac{M_s}{B_0} =\chi_0 \left[ -\Theta(-\varepsilon_0) - \sgn{(B_0)} \sum_{n=1}^{n_c} \frac{\Delta}{\sqrt{\varepsilon_c^2 n + \Delta^2}} \right] 
\end{equation}

We have assumed that the chemical potential remains pinned at zero energy and
does not change as we tune the magnetic field.  With these assumptions
the magnetization has an interesting discontinuity at a finite
magnetic field.  The discontinuity in $M_s$
occurs since tuning the magnetic field forces the fully spin polarized
Landau level energy to evolve according to Eq.\ (\ref{del1}) and cross
the chemical potential. The critical magnetic field at which the jump happens is given by
\begin{equation}\label{gapeq}
B_c =  \frac{2}{g_s \mu_B} \Delta_0.
\end{equation}
Assuming that the intrinsic gap $\Delta_0$ remains independent of
$B_0$ we plot the resulting magnetization in Fig.\ \ref{fig3}.  This
predicted behaviour could be employed to experimentally detect the
intrinsic magnetization gap in the surface of a magnetically doped TI
and measure the size of $\Delta_0$ through Eq.\
(\ref{gapeq}). Although this signature only occurs when the chemical
potential lies very close to the surface state Dirac point we note that
magnetically doped samples satisfying this requirement have been grown
and studied.\cite{MagDopped}

\begin{figure}[t]
\includegraphics[width=8cm]{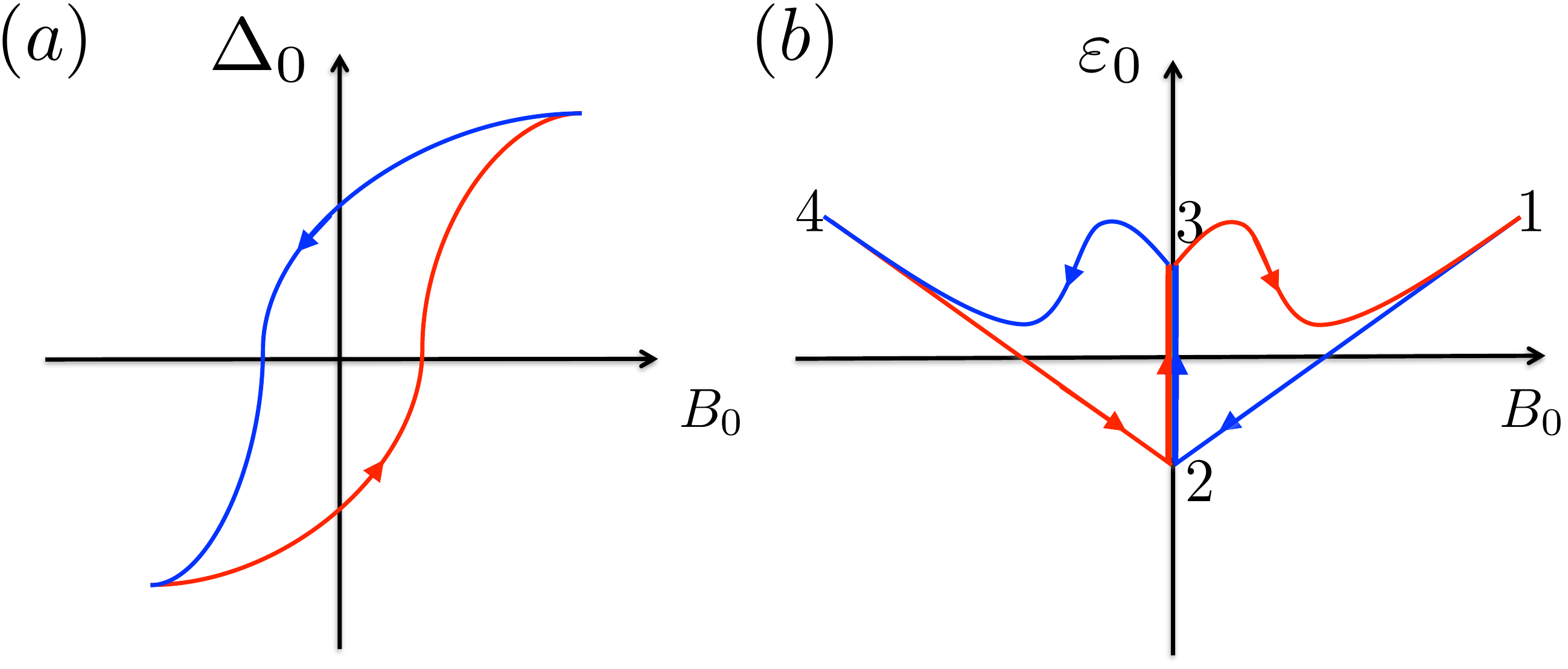} 
\caption{(Color online) (a) Schematic behaviour of $\Delta_0$, i.e., the intrinsic magnetization gap on the surface in a system with a slab geometry shown in Fig.\ \ref{fig1} versus the applied magnetic field. (b) The energy of the $n=0$ Landau level versus the 
magnetic field. Starting at the point 1 and by decreasing the magnetic field gradually, the energy would follow the path shown by the blue curve, i.e., $1\rightarrow2\rightarrow3\rightarrow4$, this happens if $\Delta_0$ is described by the blue part of the cycle in (a). Now by increasing the magnetic field from a negative value, corresponding to the point 4, the energy would follow the path given by the red curve, i.e., $4\rightarrow2\rightarrow3\rightarrow1$, since this time $\Delta_0$ would be given by the red curve in the hysteresis cycle.}\label{fig5}
\end{figure} 

It is important to note that $\Delta_0$ is in general not independent
of $B_0$ since magnetic moments of dopants will tend to align with the
applied field. Thus, like in a ferromagnet, there will be a hysteresis 
effect whose features will depend on the material details\cite{MagDopped}. 
One possible scenario is shown schematically in Fig.\ \ref{fig5}. In
the case when $\Delta_0$ depends on the field the equation
(\ref{gapeq}) continues to hold but must now be viewed as an implicit
equation for the critical field $B_c$.

Another question that arises has to do with the origin of the
electrons that fill the $n=0$ Landau level upon changing the field
through $B_c$. One may wonder where the extra electrons come from in a
fully gapped isolated system. The answer lies in the side surfaces
which under generic conditions remain gapless and act as a reservoir of electrons.
The model we consider here does not capture these states but they
reflect themselves in solutions of the Hamiltonian which are not
normalizable in an infinite system. These are fully spin polarized with opposite 
energy and spin direction. In fact they are the particle-hole
conjugates of the fully spin polarized LL given in  Eq. (\ref{phi0}).
Taking into account these electronic states and the fact that it is
more favourable for electrons with higher 
energy to be transferred to the negative energy states the counting
problem can be resolved. 

\section{The Knight Shift}
We now outline how a $\beta$-NMR experiment can in principle be used to probe
some of the physics discussed in the previous section.
The valence and conduction electrons in a metal posses magnetic
moments arising from both their 
orbital motion and their spin degrees of freedom. Nuclear
magnetic resonance technique can be used as a probe of the spin part of the 
total magnetization in the presence of the magnetic field by measuring
the relative shift in the nuclear
resonance peak with respect to the same resonance peak in a reference
insulating system. This effect, which is due to the
interaction between electronic spins and those 
of the nuclei, is known in the literature as the Knight shift and has been
extensively studied in both theory and experiment. \cite{nmrbook}
The mobile electrons in a metal interact with the nuclei in their proximity and the Knight shift in the resonance peak of these nuclei
can be described by a local Fermi contact interaction term given by $H_{\text{int}} = - {8 \pi \over 3}{\bm \mu}_e \cdot \sum_{i} {\bm \mu}_i \delta({\bm r} - {\bm R}_i)$ 
where ${\bm R}_i$ is the position of the $i$th nucleus and ${\bm \mu}_i =\gamma_N {\bm I}_i$ is its magnetic moment. The magnitude of 
$\gamma_N$, the gyromagnetic ratio, depends on the nucleus quantum state. The nucleus total spin, ${\bm I}_i$, couples to the applied 
magnetic field and therefore the position of the peak depends on the magnitude of the total magnetic field experienced by the nucleus which 
has a contribution due to the interaction with electrons. It turns out that this shift is proportional to the spin susceptibility.
The constant of proportionality, known as the hyperfine coupling, can be computed using first principle calculations for the implanted nuclei. On the other hand, 
if we assume that the presence of the nuclei does not significantly alter the electronic states, then it is possible to approximate the shift for them by taking the expectation 
value of the aforementioned interaction term using the unperturbed electronic states. This is the lowest order approximation in the perturbative 
treatment of the interaction term. For metallic systems with spin degenerate bands this shift is proportional to the spin susceptibility as it can been seen from a simple 
calculation considering the fact that the spatial and spin degrees of freedom are uncorrelated\cite{nmrbook}. 

The spin-momentum locking 
on the surface of a TI along with the energy dependence of the penetration depth can in principle
change the above simple physics. Since it is not possible anymore to
separate spin and orbital degrees of freedom, one might question the
validity of the linear relation between
the Knight shift and the spin susceptibility. We devote the rest of this section to addressing this issue by considering a very simple model. We assume that the 
nuclei do not alter the electronic states around them. It is important to note that this assumption may break down for the implanted nuclei if they 
modify the electronic states around them significantly and computing the Knight shift would then require a first principle calculations. 

The field experienced by the $i$th nucleus due to the interaction with
the proximate electrons is given by
\begin{equation}
{\bm \delta} \bB_i \equiv \ - \frac{8 \pi}{3} \gamma_e \langle{\bm \sigma} \delta(\br - \bR_i)\rangle_{T},
\end{equation}
where $\langle ... \rangle_{T}$ is the expectation value of  over electronic states at temperature $T$. Therefore, the effective Hamiltonian for the ensemble of nuclei takes the form
\begin{equation}
{ \bm H}_N^{\text{eff}} = - \hbar \gamma_N \sum_i {\bm I}_i \cdot (\bB_0 +{\bm \delta}  \bB_i).
\end{equation}
This way, $i$th nucleus would have a resonance peak $\omega_i = \gamma_N (B_0 + \delta B_{i z})$. The Knight shift is then defined by comparing 
the resonance frequency with the frequency in a similar material without these electronic states
\begin{equation}
K_i = \frac{\omega_i - \omega_0}{\omega_0} = \frac{B_0 + \delta B_{i z} - B_0}{B_0} = \frac{\delta B_{i z}}{ B_0}.
\end{equation}
The shift in the resonance peak of the nuclei ensemble is the average of the knight shift from each individual nucleus and is given by
\begin{equation}
K = \frac{1}{N} \sum_i \frac{\delta B_{i z}}{B_0},
\end{equation}
where $N$ is the number of the implanted nuclei. Using the electronic eigenstates given in Eq. (\ref{phi_p}-\ref{phi0}) we get the following expression for $K$
\begin{equation}
-\frac{8 \pi \gamma_e}{3 N B_0} \sum_i \sum^{\text{occ}}_{k_x, n, \alpha} |\psi^{\alpha}_n(z_i)|^2 (|\phi^{\alpha}_{n,\uparrow}|^2(\bR_{\bot i}) -|\phi^{\alpha}_{n,\downarrow}|^2 (\bR_{\bot i}) ).
\end{equation} 
Here $|\psi^{\alpha}_n(z)|^2$ appears as a factor in the realistic 3D electronic wave functions of the surface electrons reflecting the fact that 
the electrons have an energy dependent penetration depth into the bulk. The nuclei implanted in the system have a spatial probablity distribution 
$P_{\text{Nuc}}(z,{\bm r}_{\bot})$, which depends on the energy and diameter of the beam of the ions used in the $\beta$-NMR experiment.   
If the distribution function is known, we can replace the above summation with a 3D integral over the crystal volume 
\begin{equation}
\frac{1}{N} \sum_i \; \rightarrow \; \int dz d^2 {\bm r}_{\bot} P_{\text{Nuc}}(z,{\bm r}_{\bot}).
\end{equation}
Assuming that the distribution is uniform in the plane of the surface, i.e., $P_{\text{Nuc}}(z,{\bm r}_{\bot}) = P(z)$, we get
\begin{equation}
K =  \frac{8 \pi }{3 l_B^2 B_0}\sum^{\text{occ}}_{n,\alpha} f^{\alpha}_n  \cdot {\cal M}^{\alpha}_{z,n},
\end{equation}
where we have performed the integration over the in-plane degrees of freedom and  replaced the summation over $k_x$ with 
the LL degeneracy.  We have also defined the $n$th LL weight, $f^{\alpha}_n$, as 
\begin{equation}
f^{\alpha}_n \equiv \int dz  |\psi^{\alpha}_n(z)|^2 P(z). 
\end{equation}
Now if we assume that different LL have the same penetration depth we have $f^{\alpha}_n = f_0$ for all $n<n_c$ and we obtain
\begin{equation}
K =  \frac{8 \pi f_0}{3 B_0} \frac{1}{l_B^2}\sum^{\text{occ}}_{n,\alpha} M^{\alpha}_{z,n}   = \frac{8 \pi f_0}{3} \chi_e^s.
\end{equation}

We thus recover the linear proportionality of the Knight shift to the surface electronic spin susceptibility under reasonable assumptions. Note that if we relax the assumption that different LLs can now have different penetration depths, then the Knight shift would no longer be linearly proportional 
to the total spin susceptibility. Instead, it would be a weighted superposition of contributions from each individual LL to the spin susceptibility. Nevertheless, the Knight shift will still display the interesting behavior discussed in this study as long as $f^{\alpha}_n$ is a reasonably slowly varying function of $n$.
We should emphasize once again that although above considerations
elaborate on the differences caused by the spin-momentum locking 
and the energy dependent penetration depth, they do not take into account the fact that the electronic wave-functions could be altered by the 
presence of the implanted nuclei and the hybridization with the adjacent nuclei.
\newline
\newline
\section{Conclusions}
The magnetic response of spins of the Dirac-like electrons on the surface of a topological insulator
 shows interesting features both in the absence and the presence of an intrinsic gap $\Delta_0$. When the surface states are gapped owing to the 
time-reversal breaking perturbation (i.e.\ due to magnetic doping)
the $n=0$ Landau level, which is fully spin polarized, can have positive or negative energy depending on the sign of the intrinsic gap 
$\Delta_0$ relative to that of the applied magnetic field. It will therefore be completely filled or empty  when the chemical potential is tuned to zero energy. Our study shows that this structure results in an observable jump in the spin susceptibility, mesurable e.g. through the $\beta$-NMR Knight shift, as one tunes the applied magnetic field through the critical value $B_c$ given in Eq.\ (\ref{gapeq}). The effect may be used as a means to measure the magnitude of the intrinsic gap on the surface of a magnetically doped topological insulator if $\beta$-NMR or another surface-sensitive technique could capture the magnetic response of the electronic spins on the surface. 
This behaviour is a unique feature of the topological insulator exotic surface states closely related to the special form of the spin-momentum entanglement. 
\section{Acknowledgment} 
The authors are indebted to W.A. MacFarlane, R.F. Kiefl and K.A. Moler for illuminating discussions. This work was supported in part by NSERC and CIfAR.


\end{document}